arXiv:0806.3738v3 [q-bio.NC] 21 October 2008

# The Influence of Sodium and Potassium Dynamics on Excitability, Seizures, and the Stability of Persistent States: I. Single Neuron Dynamics.

John R. Cressman Jr.[1], Ghanim Ullah[2], Jokubas Ziburkus[3], Steven J. Schiff[2, 4], and Ernest Barreto[1]

[1] Department of Physics & Astronomy and The Krasnow Institute for Advanced Study, George Mason University, Fairfax, VA, 22030, USA, [2] Center for Neural Engineering, Department of Engineering Science and Mechanics, The Pennsylvania State University, University Park, PA, 16802, USA, [3] Department of Biology and Biochemistry, The University of Houston, Houston, TX, 77204, USA, and [4] Departments of Neurosurgery and Physics, The Pennsylvania State University, University Park, PA, 16802, USA.

## ABSTRACT

In these companion papers, we study how the interrelated dynamics of sodium and potassium affect the excitability of neurons, the occurrence of seizures, and the stability of persistent states of activity. In this first paper, we construct a mathematical model consisting of a single conductance-based neuron together with intra- and extracellular ion concentration dynamics. We formulate a reduction of this model that permits a detailed bifurcation analysis, and show that the reduced model is a reasonable approximation of the full model. We find that competition between intrinsic neuronal currents, sodium-potassium pumps, glia, and diffusion can produce very slow and large-amplitude oscillations in ion concentrations similar to what is seen physiologically in seizures. Using the reduced model, we identify the dynamical mechanisms that give rise to these phenomena. These models reveal several experimentally testable predictions. Our work emphasizes the critical role of ion concentration homeostasis in the proper functioning of neurons, and points to important fundamental processes that may underlie pathological states such as epilepsy.

## INTRODUCTION

The Hodgkin-Huxley equations (Hodgkin and Huxley, 1952) have played a vital role in our theoretical understanding of various behaviors seen in neuronal studies both at the single-cell and network levels. However, use of these equations often assumes that the intra- and extra-cellular ion concentrations of sodium and potassium are constant. While this may be a reasonable assumption for the isolated squid giant axon, its validity in other cases, especially in the mammalian brain, is subject to debate. In this first of two companion papers, we investigate the role of local fluctuations in ion concentrations in modulating the behavior of a single neuron.

Most studies investigating normal brain states have focused primarily on the intrinsic properties of neurons. Although some studies have examined the role that the extracellular micro-environment plays in pathological behavior (Bazhenov et al., 2004; Kager et al., 2000; Somjen, 2004; Park and Durand, 2006, Frohlich et al., 2008), little attention has been paid to the cellular control of micro-environmental factors as a means to modulate the neuronal response (Park and Durand, 2006).

In general, the intrinsic excitability of neuronal networks depends on the reversal potentials for various ion species. The reversal potentials in turn depend on the intra- and extracellular concentrations of the corresponding ions. During neuronal activity, the extracellular potassium and intracellular sodium concentrations ($[K]_o$ and $[Na]_i$, respectively) increase (Amzica et al., 2002; Heinemann et al., 1977; Moody et al., 1974; Ransom et al., 2000). Glia help to reestablish the normal ion concentrations, but require time to do so. Consequently, neuronal excitability is transiently modulated in a competing fashion: the local increase in $[K]_o$ raises the potassium reversal potential, increasing excitability, while the increase in $[Na]_i$ leads to a lower sodium reversal potential and thus less ability to drive sodium into the cell. The relatively small extracellular space and weak sodium conductances at normal resting potential can cause the transient changes in $[K]_o$ to have a greater effect over neuronal behavior than the changes in $[Na]_i$, and the overall increase in excitability can cause spontaneous neuronal activity (Mcbain, 1994; Rutecki et al., 1985; Traynelis and Dingledine, 1988).

In this paper, we examine the mechanisms by which the interrelated dynamics of sodium and potassium affect the excitability of neurons and the occurrence of seizure-like behavior. Since modest increases in $[K]_o$ are known to produce more excitable neurons, we seek to understand ion concentration dynamics as a possible mechanism for giving rise to and perhaps governing seizure behavior. Using the major mechanisms responsible for the upkeep of the cellular micro-environment, i.e. pumps, diffusion, glial buffering, and channels, we mathematically model a conductance-based single neuron embedded within an extracellular space and glial compartments. We formulate a reduction of this model that permits a detailed analytical bifurcation analysis of the dynamics exhibited by this model, and show that the behavior produced by the reduced model is a reasonable approximation of the full model's dynamics. The effects of ion concentration dynamics on the behavior of networks of neurons is addressed in the companion article (Ullah et al., Submitted).

Some related preliminary results have previously appeared in abstract form (Cressman et al, 2008).

**METHODS**

1. *Full model*

Our full model consists of one single-compartment conductance-based neuron containing sodium, potassium, calcium-gated potassium, and leak currents, augmented with dynamic variables representing the intracellular sodium and extracellular potassium concentrations. These ion concentrations are affected by the neuron's intrinsic ionic currents as well as a sodium-potassium pump current, a glial current, and potassium diffusion. Finally, the concentrations are coupled to the membrane voltage equations via the Nernst reversal potentials.

The conductance-based neuron is modeled as follows:

$$C\frac{dV}{dt} = I_{Na} + I_K + I_{Cl}$$

$$I_{Na} = -g_{Na}\left[m_\infty(V)\right]^3 h\left(V - V_{Na}\right) - g_{NaL}(V - V_{Na})$$

$$I_K = -\left(g_K n^4 + \frac{g_{AHP}[Ca]_i}{1+[Ca]_i}\right)(V - V_K) - g_{KL}(V - V_K)$$

$$I_{Cl} = -g_{ClL}(V - V_{Cl})$$

$$\frac{dq}{dt} = \phi[\alpha_q(V)(1-q) - \beta_q(V)q], \quad q = n, h$$

$$\frac{d[Ca]_i}{dt} = -0.002 g_{Ca}(V - V_{Ca}) / \{1 + \exp(-(V+25)/2.5)\} - [Ca]_i / 80 \qquad (1)$$

The supporting rate equations are:

$$m_\infty(V) = \alpha_m(V)/(\alpha_m(V) + \beta_m(V))$$

$$\alpha_m(V) = 0.1(V+30)/[1 - \exp(-0.1(V+30))]$$

$$\beta_m(V) = 4\exp(-(V+55)/18)$$

$$\alpha_n(V) = 0.01(V+34)/[1 - \exp(-0.1(V+34))]$$

$$\beta_n(V) = 0.125\exp(-(V+44)/80)$$

$$\alpha_h(V) = 0.07\exp(-(V+44)/20)$$

$$\beta_h(V) = 1/[1 + \exp(-0.1(V+14))]$$

Note that the overall leak current consists of the final terms in the above expressions for $I_{Na}$ and $I_K$, plus $I_{Cl}$; similar leak currents were used by (Kager et al., 2000). Also, the gating variable $m$ is assumed to be fast compared to the voltage change; we therefore assume it reaches its equilibrium value $m_\infty$ immediately (Rinzel, 1985; Pinsky and Rinzel, 1994). Finally, the active internal calcium concentration is used only in conjunction with the calcium-gated potassium current in order to model the adaptation seen in many excitatory cells (Mason and Larkman, 1990; Wang, 1998).

The meaning and values of the parameters and variables used in this paper are given in Table 1.

The potassium concentration in the interstitial volume surrounding each cell was continuously updated based on $K^+$ currents across the neuronal membrane, $Na^+$-$K^+$ pumps, uptake by the glial network surrounding the neurons, and lateral diffusion of $K^+$ within the extracellular space. Thus, we have

$$\frac{d[K]_o}{dt} = -0.33 I_K - 2\beta I_{pump} - I_{glia} - I_{diff}. \qquad (2)$$

The factor 0.33mM.cm$^2$/μcoul converts current density to rate-of-change of concentration (see Appendix A). The factor $\beta$ corrects for the volume fraction between the interior of the cell and the extracellular space when calculating the concentration change and is based on Mazel et al. (1998), McBain et al. (1990), and Somjen (2004).

The sodium-potassium pump is modeled as a product of sigmoidal functions as follows:

$$I_{pump} = \left( \frac{\rho}{1.0 + \exp((25.0 - [Na]_i)/3.0)} \right) \left( \frac{1.0}{1.0 + \exp(5.5 - [K]_o)} \right).$$

Normal resting conditions are attained when $\rho$ = 1.25mM/sec. Each term saturates for high values of internal sodium and external potassium, respectively. More biophysically realistic models of pumps, such as those in (Lauger, 1991) produce substantially similar results.

The capacity of glial cells to remove excess potassium from the extracellular space is modeled by

$$I_{glia} = \frac{G_{glia}}{1.0 + \exp((18 - [K]_o)/2.5)}.$$

This highly simplified model incorporates both passive and active uptake into a single sigmoidal response function that depends on the extracellular potassium concentration alone. Normal conditions correspond to $G_{glia}$ = 66mM/sec, and $[K]_o$ = 4.0mM. A similar but more biophysical approach was used in (Kager et al., 2000). Two factors allow the glia to provide a nearly insatiable buffer for the extracellular space. The first is the very large size of the glial network. Second, the glial endfeet surround the pericapillary space, which, through interactions with arteriole walls, can effect blood flow; this, in turn, can increase the buffering capability of the glia (Paulson and Newman, 1987, Kuschinsky et al., 1972, McCulloch et al., 1982).

The diffusion of potassium away from the local extracellular micro-environment is modeled by

$$I_{diff} = \varepsilon([K]_o - k_{o,\infty}).$$

Here, $k_{o,\infty}$ is the concentration of potassium in the largest nearby reservoir. Physiologically, this would correspond to either the bath solution in a slice preparation, or the vasculature in the intact brain (noting that $[K]_o$ is kept below the plasma level by trans-endothelial transport). For normal conditions, we use $k_{o,\infty}$ = 4.0 mM. The diffusion constant $\varepsilon$, obtained from Fick's law, is $\varepsilon = 2D/\Delta x^2$, where we use $D = 2.5\times10^{-6}$ $cm^2$/sec for $K^+$ in water (Fisher et al., 1976) and estimate $\Delta x \approx 20\mu m$ for intact brain reflecting the average distance between capillaries (Scharrer, 1944); thus $\varepsilon$ = 1.2Hz.

To complete the description of the potassium concentration dynamics, we make the assumption that the flow of $Na^+$ into the cell is compensated by flow of $K^+$ out of the cell. Then $[K]_i$ can be approximated by

$$[K]_i = 140.0mM + (18.0mM - [Na]_i), \tag{3}$$

where 140.0 mM and 18.0 mM reflect the normal resting $[K]_i$ and $[Na]_i$ respectively. The limitations of this approximation will be addressed in the discussion section.

The intra- and extracellular sodium concentration dynamics are modeled by

$$\frac{d[Na]_i}{dt} = 0.33\frac{I_{Na}}{\beta} - 3I_{pump} \tag{4}$$

$$[Na]_o = 144.0mM - \beta([Na]_i - 18.0mM). \tag{5}$$

In equation (5), we assume that the total amount of sodium is conserved, and hence only one differential equation for sodium is needed. Here, 144.0mM is the sodium concentration outside the cell under normal resting conditions for a mammalian neuron.

Finally, the reversal potentials appearing in equation (1) are obtained from the ion concentrations via the Nernst equation

$$V_{Na} = 26.64 \ln\left(\frac{[Na]_O}{[Na]_i}\right)$$

$$V_K = 26.64 \ln\left(\frac{[K]_O}{[K]_i}\right)$$

$$V_{Cl} = 26.64 \ln\left(\frac{[Cl]_i}{[Cl]_o}\right).$$

With the leak conductances listed above, the chloride concentrations were fixed at $[Cl]_i$=6.0mM and $[Cl]_o$=130mM.

Thus, the dynamic variables of the full model are V, n, h, $[Ca]_i$, $[K]_o$, and $[Na]_i$. Despite the fact that we have neglected many features of real mammalian cells (such as the geometrically complex dendritic and axonal structure and the related spatially complex distribution of channels and cotransporters, as well as the presence of immobile anions which are strictly required to maintain electric and osmotic balance), our model captures the essential dynamics that we wish to explore.

In the results section, we will be interested in varying the parameters $G_{glia}$, $\varepsilon$, $k_{o,\infty}$, and $\rho$. We will present our results in terms of parameters normalized by their normal values, for example, $\bar{G}_{glia} = G_{glia} / G_{glia,normal}$, where the overbar indicates the normalized parameter.

2. *Reduced model*

In order to more effectively study the bifurcation structure of the model presented above, we formulated a reduction by eliminating the fast-time-scale spiking behavior in favor of the slower ion concentration dynamics. This is accomplished by replacing the entire Hodgkin-Huxley mechanism with empirical fits to time-averaged ion currents.

Using the membrane conductances from the full model, we fixed the internal and external sodium and potassium concentration ratios and allowed the model cell to attain its asymptotic dynamical state, which was either a resting state or a spiking state. Then, the sodium and potassium membrane currents were time-averaged over one second. These data were fit to products of sigmoidal functions of the sodium and potassium concentration ratios, resulting in the (infinite-time) functions $I_{Na\infty}\left([Na]_i/[Na]_o,[K]_o/[K]_i\right)$ and $I_{K\infty}\left([Na]_i/[Na]_o,[K]_o/[K]_i\right)$. Details are available in Appendix A. $I_{Na\infty}$ is nearly identical to $I_{K\infty}$, differing significantly only near normal resting concentration ratios due to differences in the sodium and potassium leak currents.

Thus, our reduced model consists of equations (2-5), with $I_{Na}$ and $I_K$ replaced with the empirical fits described above (see Appendix A for additional details).

*3. Bifurcations in the reduced model*

Our main results in this paper consist of identifying bifurcations in the reduced model and analyzing their implications for the behavior of the full model. We observe that depending on the various parameters, the ion concentrations in the reduced model approach either stable *equilibria*, and thus remain constant, or they approach stable *periodic orbits*, and thus exhibit oscillatory behavior (Fig 1). As parameters are changed, the stability of these solutions change. This happens through *bifurcations* (a good general reference is Strogatz (1994)). Most relevant for our purposes are the *Hopf bifurcation*, and the *saddle-node bifurcation of periodic orbits*. In a Hopf bifurcation, an equilibrium solution either gains or loses stability, and simultaneously, a periodic orbit either appears or disappears from the same point[1]. In a saddle-node bifurcation of periodic orbits, a pair of periodic orbits – one stable and one unstable – either appears or disappears suddenly, as if out of “thin air”.

Bifurcation diagrams were obtained using XPPAUT (Ermentrout, 2002). Code for both of our models is available from ModelDB.[2]

[1] Depending on the stabilities of the equilibrium and the periodic orbit involved, Hopf bifurcations are classified as sub- or supercritical.
[2] http://senselab.med.yale.edu/modeldb/

**RESULTS**

1. *Overview*

In an experimental slice preparation, an easily-performed experimental manipulation is to change the potassium concentration in the bathing solution. Such preparations have been used to study epilepsy (Jensen and Yaari, 1997; Traynelis and Dingledine, 1988; Gluckman, et al. 2001). At normal concentrations (~4mM), normal resting potential is maintained. However, at higher concentrations (8mM, for example) bursts and seizure-like events occur spontaneously.

We begin discussing the dynamics of our models by considering a similar manipulation, corresponding to varying the normalized parameter $\bar{k}_{o,\infty}$. In the full model, setting $\bar{k}_{o,\infty}$ = 2.0 (i.e., doubling the normal concentration of potassium in the bath solution) leads to spontaneously-occurring prolonged periods of rapid firing, as illustrated in the top trace of Fig. 1. These oscillations are remarkably similar to experimental results reported by several investigators (see, for example, Figures 1 and 6 of Jensen and Yaari (1997), in which the authors use a high potassium *in vitro* preparation, and Figure 2 of Ziburkus et al. (2006), which reports results from a 4-aminopyridine *in vitro* preparation). Each event lasts on the order of tens of seconds and consists of many spikes, each of which occurs on the order of 1 ms. Thus, the full model contains dynamics on at least two distinct time scales that are separated by four orders of magnitude: fast spiking from the Hodgkin-Huxley mechanism, and a slow overall modulation. The solid traces in the middle and bottom panels show that this slow modulation corresponds to slow periodic behavior in the sodium and potassium ion concentrations, respectively.

Our reduced model was constructed in order to remove the fast Hodgkin-Huxley spiking mechanism and focus attention on the slow dynamics of the ion concentrations. The dashed traces in the middle and bottom panels of Fig. 1 show the sodium and potassium ion concentrations obtained from the reduced model for the same parameters used above. Although these traces are not identical to those of the full model, it is evident that the reduced model captures the qualitative behavior of the ion concentrations quite well.

The separation of time scales achieved by our model reduction (see, for example, Rinzel and Ermentrout (1989); Kepler, et al., (1992)) yields a model that is amenable to numerical bifurcation analysis. Knowledge of the bifurcation structure in turn informs us about the dynamical mechanisms that underlie the full model. In the following, we will first describe the main dynamical features of the reduced model, and then examine the implications for the behavior of the full model.

2. *Analysis of the reduced model*

Fig. 2 shows a bifurcation diagram obtained using the reduced model. This diagram plots the minimum and maximum asymptotic values of the extracellular potassium concentration $[K]_o$ versus a range of values of the reservoir's normalized potassium concentration $\bar{k}_{o,\infty}$. For low values of this parameter, $[K]_o$ is observed to settle at a stable equilibrium. The value of $[K]_o$ corresponding to this equilibrium increases with $\bar{k}_{o,\infty}$ until the equilibrium loses stability via a subcritical Hopf bifurcation at $\bar{k}_{o,\infty} \approx 1.9$. This means that at this point, an unstable periodic orbit collapses onto the equilibrium, and both disappear. $[K]_o$ is subsequently attracted to a coexisting large-amplitude stable periodic orbit[3]. Thus, large-amplitude oscillations in $[K]_o$ appear abruptly. These oscillations persist as $\bar{k}_{o,\infty}$ is increased until the same sequence of bifurcations occurs in the opposite order at $\bar{k}_{o,\infty} \approx 2.13$. At this higher value of $\bar{k}_{o,\infty}$, the unstable equilibrium undergoes a subcritical Hopf bifurcation, becoming stable and giving rise to an unstable periodic orbit whose amplitude quickly rises with increasing $\bar{k}_{o,\infty}$. This orbit then collides with the large-amplitude stable periodic orbit at $\bar{k}_{o,\infty} \approx 2.15$, and both orbits disappear in a saddle-node bifurcation of periodic orbits. In this manner, the periodic behavior of $[K]_o$ is terminated. For still higher values of $\bar{k}_{o,\infty}$, $[K]_o$ approaches the equilibrium values shown at the far right in Fig. 2.

[3] The stable and unstable periodic orbits involved in this scenario appear via a saddle-node bifurcation at a slightly smaller parameter value that is extremely close to that of the Hopf bifurcation. Thus, the sequence of bifurcations is not immediately apparent in Fig. 2. The abruptness of these transitions, and the difficulty in resolving them numerically, is due to the "canard" mechanism (Dumortier and Roussarie, 1996; Wechselberger, (2007)).

In order to examine the boundaries of the oscillatory behavior described above with respect to pump strength $\bar{\rho}$, the diffusion coefficient $\bar{\varepsilon}$, and glial buffering strength $\bar{G}_{glia}$, we constructed the bifurcation diagrams shown in Fig. 3. First, we fixed all parameters at their normal resting values except $\bar{k}_{o,\infty}$, which we set to 2 in order to obtain the oscillatory behavior discussed above. We then separately varied $\bar{\rho}$, $\bar{\varepsilon}$, and $\bar{G}_{glia}$ away from their normal resting values. If $\bar{\rho}$ and $\bar{\varepsilon}$ are increased from their nominal values of 1 (Fig. 3a, b), we see that the oscillatory behavior terminates in a manner similar to that described above; that is, an unstable periodic orbit appears via a subcritical Hopf bifurcation which grows until it collides with and annihilates the stable periodic solution at a saddle-node bifurcation of periodic orbits. (This is most apparent in Fig 3(a).) The same scenario applies as these parameters are decreased[4],[5]. The situation is similar for the glial strength $\bar{G}_{glia}$, except that on the left, we see no saddle-node bifurcation of periodic orbits for positive values of $\bar{G}_{glia}$ (Fig. 3c).

It is notable that if $\bar{\varepsilon}$ or $\bar{G}_{glia}$ are reduced from their normal values, larger amplitude oscillations in $[K]_o$ occur. This is because, in both cases, the trafficking of potassium away from the extracellular space is impeded, and consequently, $[K]_o$ builds up more effectively during the spiking phase of the cell's activity (see Fig. 6, below). In contrast, changing $\bar{\rho}$ in either direction results in very little change in the amplitude of the $[K]_o$ oscillations. Furthermore, the bistable region on the right side of the $\bar{\rho}$ diagram is quite wide, and hence hysteretic behavior as $\bar{\rho}$ is varied across this region may be particularly amenable to experimental observation (e.g. with ouabain). Finally, we note that the right branch of stable equilibria in Fig. 3(a) corresponds to higher values of $[K]_o$ than the left branch. This is because the cell is active – either spiking or in depolarization block – and thus there is a large membrane potassium current flowing into the extracellular space which must be balanced by pumps and otherwise "normal" diffusion and glial currents.

---

[4] A canard similar to that described previously occurs here, so that the Hopf and the saddle-node bifurcations on the left sides of Figs. 3 (a) and (b) occur in extremely narrow intervals of the parameter.

[5] In Fig. 3 (a), the equilibrium curve does not extend all the way to zero because of the constant chloride leak current.

The two-parameter bifurcation diagram shown in Fig. 4 provides a more complete understanding of the oscillatory behavior of $[K]_o$ in our reduced model with respect to the variation of both $\bar{\varepsilon}$ and $\bar{G}_{glia}$, with $\bar{k}_{o,\infty} = 2$. The dashed lines at $\bar{G}_{glia} = 1$ and $\bar{\varepsilon} = 1$ correspond to the one-dimensional bifurcation diagrams shown in Figs. 3b and c. The solid curves represent Hopf bifurcations, and the intersection of these dashed lines with the Hopf curves correspond to the Hopf bifurcations (points) in the earlier figures. Thus, the Hopf curves define a region within which $[K]_o$ is obliged to oscillate, because the only stable attractor is a periodic orbit. To facilitate discussion, we refer to this region as the "region of oscillation", or RO. Outside of the RO, stable equilibrium solutions for $[K]_o$ exist[6].

The dashed line at $\bar{G}_{glia} = 1.75$ corresponds to the one-dimensional bifurcation diagram shown in Fig. 5. This is drawn at the same scale as Fig. 3b (the $\bar{G}_{glia} = 1$ diagram) to facilitate comparison. We note that the amplitude of the oscillation in $[K]_o$ is significantly smaller in this region of the RO. Furthermore, the Hopf bifurcation on the right (at about $\bar{\varepsilon} = 3.2$) is now supercritical. This means that the amplitude of the $[K]_o$ oscillation decays smoothly to zero as this point is approached from the left.

## 3. *Analysis of the full model*

We now investigate whether the dynamical features identified above in our reduced model correspond to similar features in our full model. Fig. 6 shows traces of the membrane voltage (upper traces), $[K]_o$ (solid lower traces), and $[Na]_i$ (dashed lower traces) versus time, all obtained from the full model, corresponding to the parameter values marked by the numbered squares in Fig. 4. For the regions outside of the RO, the reduced model predicts stable equilibrium solutions for the ion concentrations. For example, at point 1, $[K]_o$ is slightly elevated at a value near 6mM, and the membrane voltage of the full model remains constant at -62 mV (not shown). However, at point 2, the full model exhibits tonic firing, as shown in Fig. 6a. Here, $[K]_o$ is sufficiently high such that the neuron is depolarized beyond its firing threshold, and $[K]_o$ remains essentially constant with only small perturbations of order 0.1 mM due to individual spikes (Fig. 6a, lower panel) (Frankenhaeuser and Hodgkin, 1956). (These spike perturbations disappear in the reduced model, as they are smoothed-out by the averaging in our model reduction procedure.)

[6] Note that oscillations may persist slightly outside of the RO, where a stable periodic orbit coexists with the stable equilibrium solution; see, for example, the right side of Fig. 3(a).

Within the RO, the reduced model predicts periodic behavior with relatively large and slow oscillations in the ion concentrations. Points 3-7 in Fig. 4 correspond to Figs. 6 (b-f), which show various bursting behaviors of the full model. To facilitate comparison, the time scale for these figures (Fig. 6 b-f) is the same, showing 100 seconds of data. In addition, the voltage and concentration scales are also the same, except for the concentration scale in (b). We make the following observations from Fig. 6. As the RO is traversed from low to high $\bar{\varepsilon}$ (keeping $\bar{G}_{glia}$ fixed) in Fig. 4, the bursts become more frequent (compare Fig. 6c, d; points 4, 5 and Fig.6e, f; points 6, 7). In addition, the shape of the burst envelope changes due to the decreasing amplitude of the $[K]_o$ oscillations. Note also in Fig. 6b (point 3) that the peak of the $[K]_o$ concentration is nearly 40mM, large enough to cause the neuron to briefly enter a state of quiescence known as "depolarization block" (see companion paper, Ullah, et al., Submitted). Finally, the within-burst spike frequency is essentially constant in Figs. 6 (b)-(f); it does peak in concert with the peak of $[K]_o$, however, this effect is weak.

Returning briefly to the reduced model, we address in Fig. 7 how the location of the RO changes with $\bar{k}_{o,\infty}$ as all other parameters are kept at their normal values. As $\bar{k}_{o,\infty}$ is increased from its normal value at 1, the RO emerges around a value of $\bar{k}_{o,\infty} = 1.77$, as represented by the gray line on the left of Fig.7a. (This observation is consistent with Rutecki , et al. (1985)). As $\bar{k}_{o,\infty}$ increases, the right edge of the RO shifts towards the right, crossing the normal values of $\bar{\varepsilon}$ and $\bar{G}_{glia}$ at (1,1) when $\bar{k}_{o,\infty}$ = 1.9, as shown by the thick solid line in Fig.7a. As $\bar{k}_{o,\infty}$ is further increased to 2.0, normal conditions for glial pumping and diffusion are well inside the RO as shown in Fig. 7b, and correspondingly, we observe bursting/seizing behavior in the full model. Fig. 7c shows the RO when the reservoir concentration has been increased to 2.1. Here the left side edge of the RO is just about to cross the point (1,1), after which the ion concentrations assume stable equilibrium values in the reduced model. Note that these stable equilibria actually correspond to a state of rapid tonic firing in the full model, as in Fig. 6(a); for much higher values of $\bar{k}_{o,\infty}$ (greater than approximately 7.2mM), $[K]_o$ remains constant in the reduced model, but the full model eventually gives rise to depolarization block.

We conclude by reporting other kinds of bursting behaviors seen in our full model. Fig. 8a shows a time trace of the membrane voltage for $\bar{k}_{o,\infty} = 6$, $\bar{G}_{glia} = 0.1$ and $\bar{\varepsilon} = 0.4$. These bursts are fundamentally different from those shown in Fig. 6. In particular, the extracellular potassium concentration is quite elevated, and thus the periods of quiescence correspond not to resting behavior, but rather to a state of depolarization block. In addition, the bursts themselves have a rounded envelope, as opposed to the (approximately) square envelopes of the events shown in Fig. 6. This behavior is consistent with the experimental observations of (Ziburkus et al., 2006), in which interneurons were seen to enter depolarization block and thus give way to pyramidal cell bursts. Bikson et al. also observed depolarization block in pyramidal cells during electrographic seizures (see Figure 1d from Bikson et al., 2003). We have also experimentally observed (in oriens interneurons exposed to 4-aminopyridine) relatively continuous "burst" firing without any quiescent intervals, as seen in Fig.8b. In this figure, the neuron fires continuously, but with a wavy envelope due to the oscillating ion concentrations. We include these patterns to complete the description of the repertoire of single cell bursting behaviors seen in our models.

**DISCUSSION**

We have created a model which can be used to investigate the role of ion concentration dynamics in neuronal function, as well as a reduced model which is amenable to bifurcation analysis. Such bifurcations indicate major qualitative changes in system behavior, which are in many ways more predictive and informative than purely quantitative measurements. In particular, we show that under otherwise normal conditions, there exists a broad range of bath potassium concentration values which yields seizure-like behavior in a single neuron that is both qualitatively as well as quantitatively similar to what is seen in experimental models (Ziburkus et al., 2006; Feng and Durand, 2006; McBain, 1994). In fact, the values of extracellular concentration used in those experiments are quantitatively consistent with the range of concentrations shown here to exhibit seizures. Furthermore, the stable periodic oscillations in the extracellular potassium concentration which result from varying various experimentally and biophysically relevant parameters suggests that these effects may be an important mechanism underlying epileptic seizures.

In formulating our models, we made several approximations. The two most severe, insatiable glial buffering and the assumption that internal potassium can be calculated using equation (3), should hold for times that are long compared to the time scale of individual spikes, bursts, and seizures. However, for even longer times (on the order of thousands of seconds), the saturation of glia as well as the decoupling of the internal potassium from the sodium dynamics will lead to more substantial errors in the calculated results. It is possible that glia do not fully saturate, if, as suggested by Paulson and Newman et al (1987), the glia siphon excess potassium into pericapillary spaces via the astrocyte network. Nevertheless, one can understand a slow partial saturation of the glial network as a slow decrease in $\bar{G}_{glia}$, for example. Consequently, the system may enter or leave parameter-space regions in which oscillating ion concentrations exist (e.g., see Figs. 3(c) and 4). This long-term behavior could be used to more accurately model the temporal dynamics of the glial siphoning system.

It is important to note the limitations of our models with respect to extremely long time scales. If the reservoir concentration $\bar{k}_{o,\infty}$ remains only slightly elevated for a long period of time the model cell will ultimately reach a new fixed point in $[K]_o$ nearly equal to the bath concentration. A stable resting state should exhibit some degree of robustness in its micro-environment. However, as the system drifts further from the normal state, we should not expect such homeostasis to persist; the internal potassium will in general drift to higher or lower values depending on the wide variety of pumps, cotransporters, or channels inherent to the cell. When the internal potassium is integrated separately (not shown here), we see both upward and downward drift depending on model parameters as well as the initial conditions for the ionic concentrations. Therefore in our model the seizure-like events, as well as the tonic firing reported here, appear to be transients on extremely long time scales. Of course, such phenomena are also transients on long time scales in animals and humans.

Although our reduced model does a good job reproducing the qualitative results of our full spiking model, there are regions were the two models disagree. These two models produce very good agreement in the region of the two-parameter graph presented in Figure 4. However the reduced model predicts the cessation of all oscillations as $\bar{G}_{glia}$ is increased past a value of 4 (not shown), whereas the full model exhibits burst-like oscillations for far greater values. This discrepancy is due to the use of relatively simple functions used in our fitting of the time-averaged Hodgkin-Huxley currents (see Appendix A). A more sophisticated fit of these data would improve the agreement between our two models.

Our work points out the important role that ion concentration dynamics may play in understanding neuronal dynamics, including pathological dynamics such as seizures. Of course, in realistic situations, these are due to a combination of local environmental conditions and electrical and chemical communication between cells (see the accompanying paper, Ullah, et al., Submitted). The models presented here, however, demonstrate that recurring seizure-like events can occur *in a single cell* that is subject to intra- and extracellular ion concentration dynamics (see also discussion in (Kager, et al., 2000, Kager, et al., 2007) regarding single-cell seizure dynamics). In addition, we have identified the basic mechanism that can give rise to such events: Hopf bifurcations that lead to slow oscillations in the ion concentrations.

**ACKNOWLEDGEMENTS**

This work was funded by NIH Grants K02MH01493 (SJS), R01MH50006 (SJS, GU), F32NS051072 (JRC), and CRCNS-R01MH079502 (EB).

## FIGURE LEGENDS

**Fig. 1**: Comparison of the reduced model to the full spiking model. The top plot shows the membrane voltage for the neuron in the full model. The middle traces show $[K]_o$ for both the full model (solid line) and the reduced model (dashed line). The bottom traces show $[Na]_i$ with the same convention. All data were integrated with an elevated bath potassium concentration at $\bar{k}_{o,\infty} = 2.0$, with all other parameters set to their normal values.

**Fig. 2**: The bifurcation diagram for $[K]_o$ as a function of the bath potassium concentration $\bar{k}_{o,\infty}$, revealing a region of oscillatory behavior. All other parameters were set equal to their normal values. Triangles represent equilibria (i.e., steady states), and circles represent periodic orbits (i.e., oscillatory behavior); stability is denoted by solid (stable) and open (unstable) symbols.

**Fig. 3**: Bifurcation diagrams for $[K]_o$ as a function of (a) the normalized pump strength, (b) the diffusion coefficient, and (c) the glial strength. All plots were produced with an elevated bath potassium concentration at $\bar{k}_{o,\infty} = 2.0$.

**Fig. 4**: A two-dimensional bifurcation diagram showing the boundaries of the region of oscillation (RO) as a function of the diffusion coefficient and the glial strength. The black curves denote Hopf bifurcations; within this region, the ion concentrations exhibit oscillatory behavior. The dashed lines correspond to the one-dimensional bifurcation diagrams in Figures 3b, 3c and 5 (see text). Examples of the dynamics of the full model, obtained at parameter values corresponding to the numbered points, are shown in Figure 6.

**Fig. 5**: The one-dimensional bifurcation diagram for $[K]_o$ as a function of the normalized diffusion coefficient $\bar{\varepsilon}$ for $\bar{k}_{o,\infty} = 2.0$ and $\bar{G}_{glia} = 1.75$.

**Fig. 6**: Examples of the dynamics of the full model, obtained at parameter values corresponding to the numbered points in Figure 4. The top trace shows the membrane voltage and the lower traces show $[K]_o$ (solid trace) and $[Na]_i$ (dashed trace) on the same time scale.

**Fig. 7**: The effect of changing the bath concentration on the location of the region of oscillation (RO) is illustrated for $\bar{\rho} = 1$ and various values of the bath potassium concentration $\bar{k}_{o,\infty}$. The square represents normal values of the diffusion and glial strength. In (a), the RO is seen to appear and move to the right as the bath potassium concentration is increased from $\bar{k}_{o,\infty} = 1.77$ (grey curve) to $\bar{k}_{o,\infty} = 1.9$ (black curve), where it intersects the square (compare the left bifurcation in Figure 2). In (b), the normal square lies within the RO for $\bar{k}_{o,\infty} = 2.0$. In (c), $\bar{k}_{o,\infty} = 2.1$, the RO has moved further to the right, and the square is close to the left boundary.

**Fig. 8**: Other bursting patterns. Traces similar to those in Figure 6 obtained with the full model for (a) $\bar{k}_{o,\infty} = 6.0$, $\bar{G}_{glia} = 0.1$, and $\bar{\varepsilon} = 0.4$; and (b) same, but with $\bar{k}_{o,\infty}$ reduced slightly. The quiescent states in (a) correspond to depolarization block; see text for further description.

Table 1: Model variables and parameters.

| Variable | Units | Description |
| --- | --- | --- |
| $V$ | mV | Membrane potential |
| $I_{Na}$ | μA/cm$^2$ | Sodium current |
| $I_K$ | μA/cm$^2$ | Potassium current |
| $I_L$ | μA/cm$^2$ | Leak current |
| $m_\infty(V)$ | | Activating sodium gate |
| $h$ | | Inactivating sodium gate |
| $n$ | | Activating potassium gate |
| $\alpha(V)$ | | Forward rate constant for transition between open and close state of a gate |
| $\beta(V)$ | | Backward rate constant for transition between open and close state of a gate |
| $[Ca]_i$ | mM | Intracellular calcium concentration |
| $V_{Na}$ | mV | Reversal potential of persistent sodium current |
| $V_K$ | mV | Reversal potential of potassium current |
| $[Na]_o$ | mM | Extracellular sodium concentration |
| $[Na]_i$ | mM | Intracellular sodium concentration |
| $[K]_o$ | mM | Extracellular potassium concentration |
| $[K]_i$ | mM | Intracellular potassium concentration |
| $I_{pump}$ | mM/sec | Pump current |
| $I_{diff}$ | mM/sec | Potassium diffusion to the nearby reservoir |
| $I_{glia}$ | mM/sec | Glial uptake |
| **Parameter** | **Value** | **Description** |
| $C$ | 1μF/cm$^2$ | Membrane capacitance |
| $g_{Na}$ | 100mS/m$^2$ | Conductance of persistent sodium current |
| $g_K$ | 40mS/m$^2$ | Conductance of potassium current |
| $g_{AHP}$ | 0.01mS/m$^2$ | Conductance of afterhyperpolarization current |
| $g_{KL}$ | 0.05mS/m$^2$ | Conductance of potassium leak current |
| $g_{NaL}$ | 0.0175mS/m$^2$ | Conductance of sodium leak current |
| $g_{ClL}$ | 0.05mS/m$^2$ | Conductance of chloride leak current |
| $\phi$ | 3sec$^{-1}$ | Time constant of gating variables |
| $V_{Cl}$ | -81.93mV | Reversal potential of chloride current |
| $g_{Ca}$ | 0.1mS/m$^2$ | Calcium conductance |
| $V_{Ca}$ | 120mV | Reversal potential of calcium |
| $\beta$ | 7.0 | Ratio of intracellular to extracellular volume of the cell |
| $\rho$ | 1.25mM/sec | Pump strength |
| $G_{glia}$ | 66mM/sec | Strength of glial uptake |
| $\varepsilon$ | 1.2sec$^{-1}$ | Diffusion constant |
| $k_{o,\infty}$ | 4.0mM | Steady state extracellular concentration |
| $[Cl]_i$ | 6.0mM | Intracellular chloride concentration |
| $[Cl]_o$ | 130.0mM | Extracellular chloride concentration |

## APPENDIX A

*Current to concentration conversion:*

In order to derive the ion concentration dynamics, we begin with the assumption that the ratio of the intracellular volume to the extracellular volume is $\beta$ = 7.0 (Somjen, 2004). This corresponds to a cell with intracellular and extracellular space of 87.5% and 12.5% of the total volume respectively. For the currents across the membrane, conservation of ions requires

$$\Delta c_i Vol_i = -\Delta c_o Vol_o,$$

where $c$ and $Vol$ represent ion concentration and volume respectively, $\Delta$ indicates change, and the subscripts *i, o* correspond to the intra- and extracellular volumes. The above equation leads to

$$\Delta c_i = -\Delta c_o \left( \frac{Vol_o}{Vol_i} \right) = -\frac{\Delta c_o}{\beta}.$$

Let $I$ be the current density in units of μA/cm$^2$ from the Hodgkin-Huxley model. Then, the total current $i_{total} = IA$ entering the intracellular volume produces a flow of charge equal to $\Delta Q = i_{total}\Delta t$ in a time $\Delta t$, where $A$ is membrane area. The number of ions entering the volume in this time is therefore $\Delta N = i_{total}\Delta t / q$ where $q$ is 1.6x10$^{-19}$ coul. The change in concentration $\Delta c_i = \Delta N / N_A Vol_i$ depends on the volume of the region to which the ions flow, where Avogadro's number $N_A$ converts the concentration to molars. The rate of change of concentration, or concentration current $dc_i / dt = i_{c,i}$, is related to the ratio of the surface area of the cell to the volume of the cell as follows

$$i_{c,i} = \frac{\Delta c_i}{\Delta t} = \frac{\Delta N}{\Delta t Vol_i N_A} = \frac{i_{total}}{q Vol_i N_A} = \frac{IA}{q Vol_i N_A} = \frac{I}{\alpha}.$$

For a sphere of radius 7μm, $\alpha$ = 21mcoul/M.cm$^2$. An increase in cell volume would result in a smaller time constant and therefore slower dynamics.

For the outward current the external ion concentration is therefore given as

$$i_{c,o} = \beta i_{c,i} = \frac{\beta I}{\alpha} = 0.33I$$

*Equations for reduced model:*

The reduced model uses empirical fits of the average membrane currents of the Hodgkin-Huxley model neuron, as described in the main text. The fits are given below.

$$I_{K\infty} = \alpha_K (g_1 g_2 g_3 + \mathrm{g_{lK}})$$

$$I_{Na\infty} = \alpha_{Na} (g_1 g_2 g_3 + \mathrm{g_{lNa}})$$

$$K_{o/i} = [K]_o / [K]_i$$

$$Na_{i/o} = [Na]_i / [Na]_o$$

$$g_1 = 420.0(1\text{-}\mathrm{A_1}(1\text{-}\mathrm{B_1}\exp(\text{-}\mu_1 \mathrm{Na_{i/o}}))^{1/3})$$

$$g_2 = \exp(\sigma_2(1.0\text{-}\lambda_2 \mathrm{K_{o/i}})/(1.0+\exp(\text{-}\mu_2 \mathrm{Na_{i/o}})))$$

$$g_3 = (1/(1+\exp(\sigma_3(1.0+\mu_3 \mathrm{Na_{i/o}}\text{-}\lambda_3 \mathrm{K_{o/i}})))^5)$$

$$g_4 = (1/(1+\exp(\sigma_4(1.0+\mu_4 \mathrm{Na_{i/o}}\text{-}\lambda_4 \mathrm{K_{o/i}})))^5)$$

$$\mathrm{g_{lK}} = \mathrm{A_{lK}}\exp(\text{-}\lambda_{\mathrm{lK}} \mathrm{K_{o/i}})$$

$$\mathrm{g_{lNa}} = \mathrm{A_{lNa}}$$

where

$\alpha_K = 1.0, \alpha_{Na} = 1.0, A_1 = .75, B_1 = .93, \mu_1 = 2.6, \lambda_2 = 7.41,$
$\sigma_2 = 2.0, \mu_2 = 2.6, \sigma_3 = 35.7,\ \mu_3 = 1.94, \lambda_3 = 24.3, \sigma_4 = .88,$
$\mu_4 = 1.48, \lambda_4 = 24.6, \mathrm{A_{lNa}} = 1.5, \mathrm{A_{lK}} = 2.6, \lambda_{lK} = 32.5$

**FIGURES**

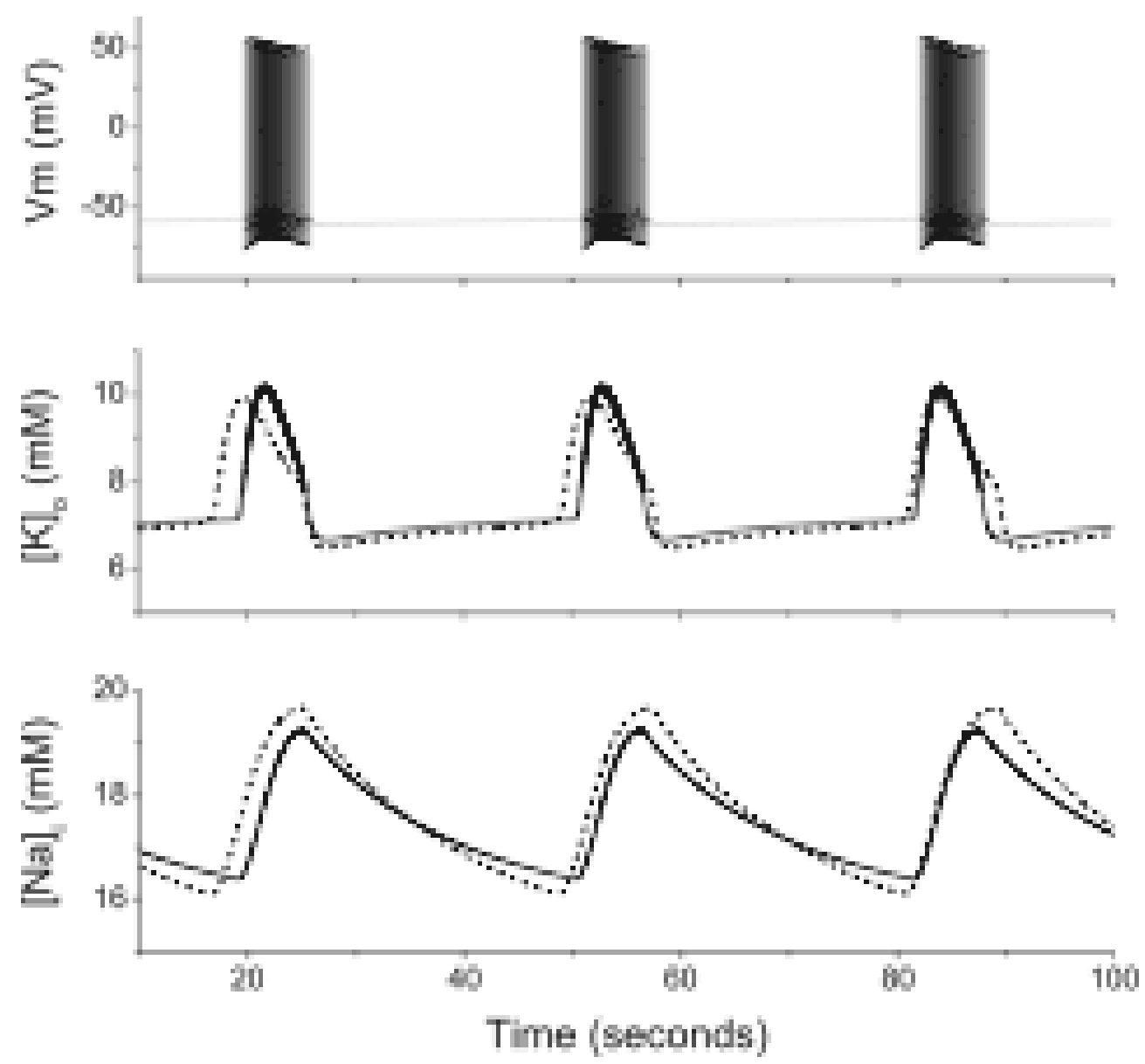

Fig. 1

**FIGURES**

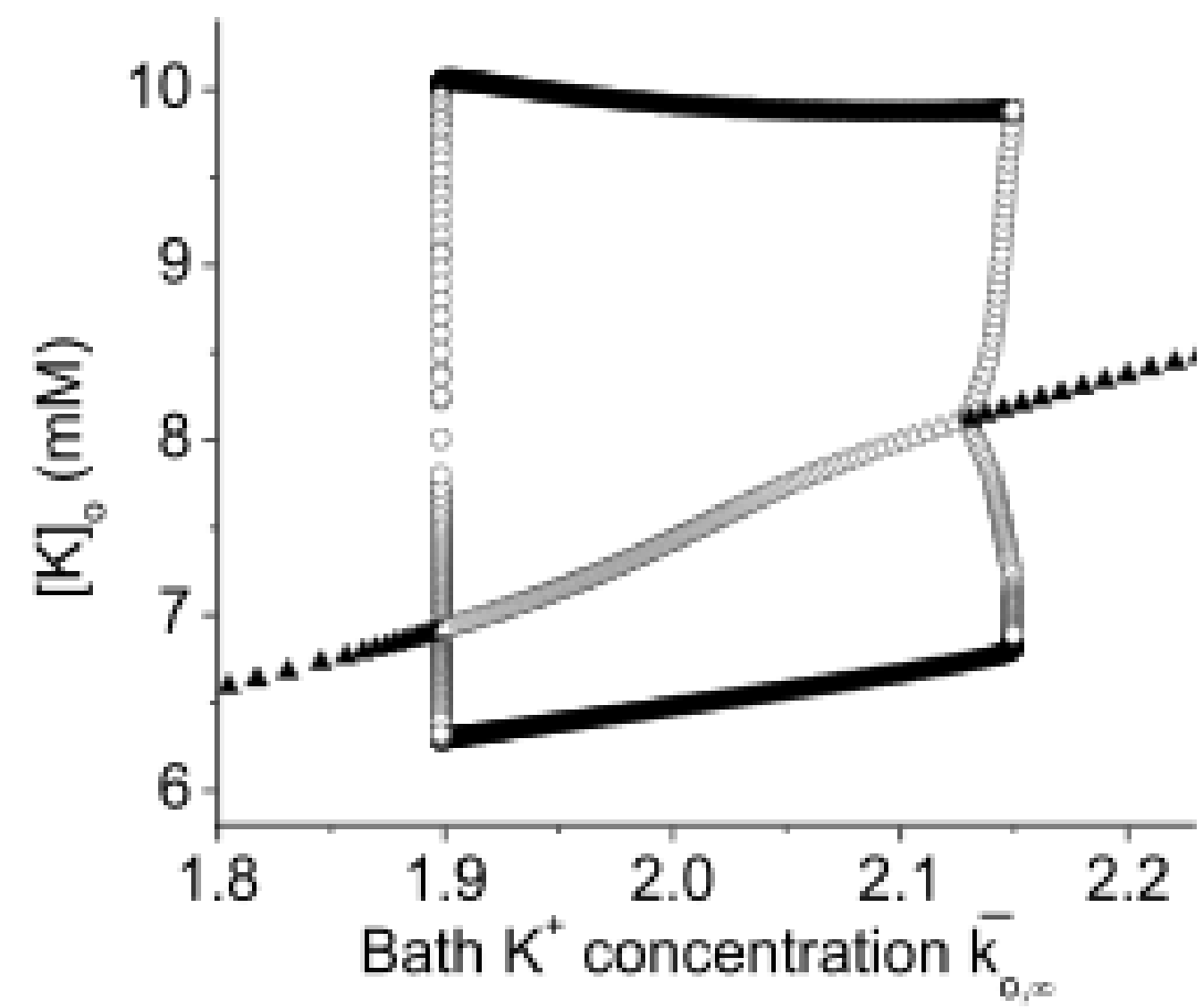

Fig. 2

**FIGURES**

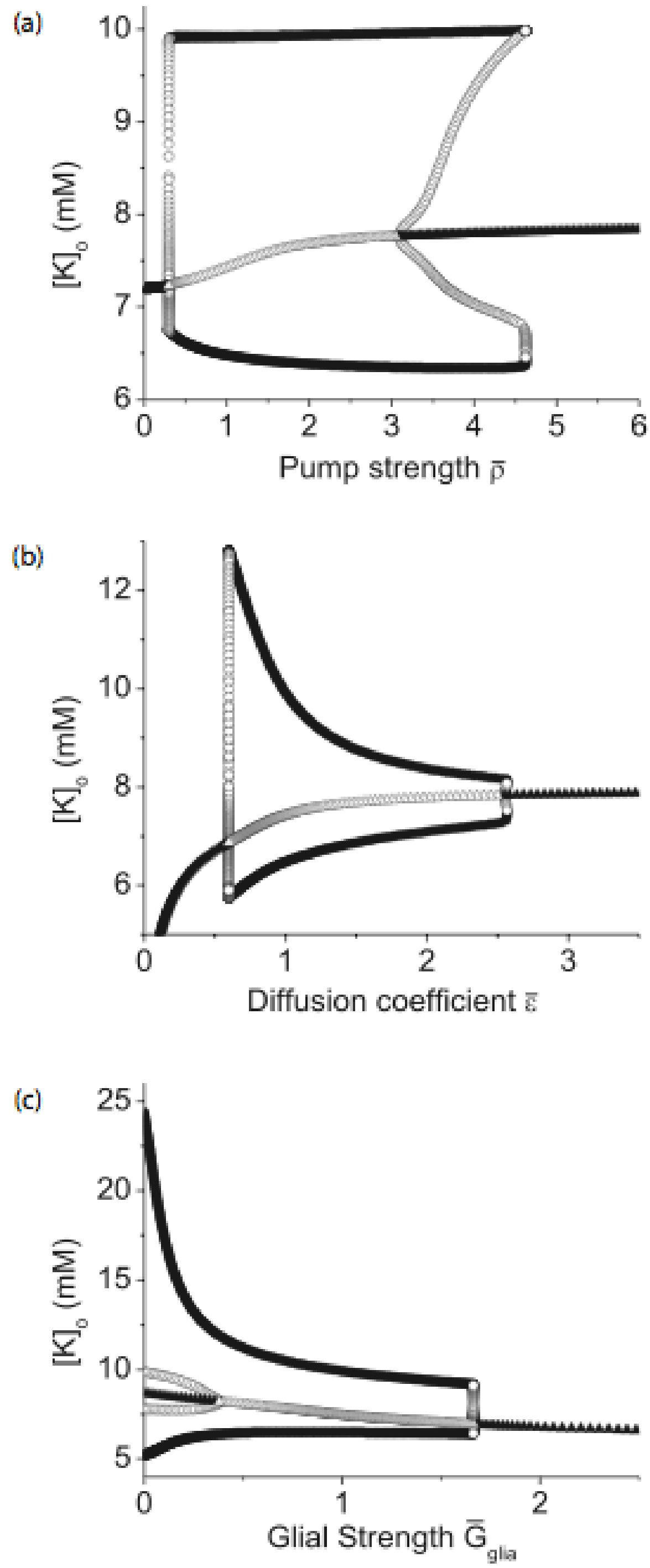

Fig. 3

**FIGURES**

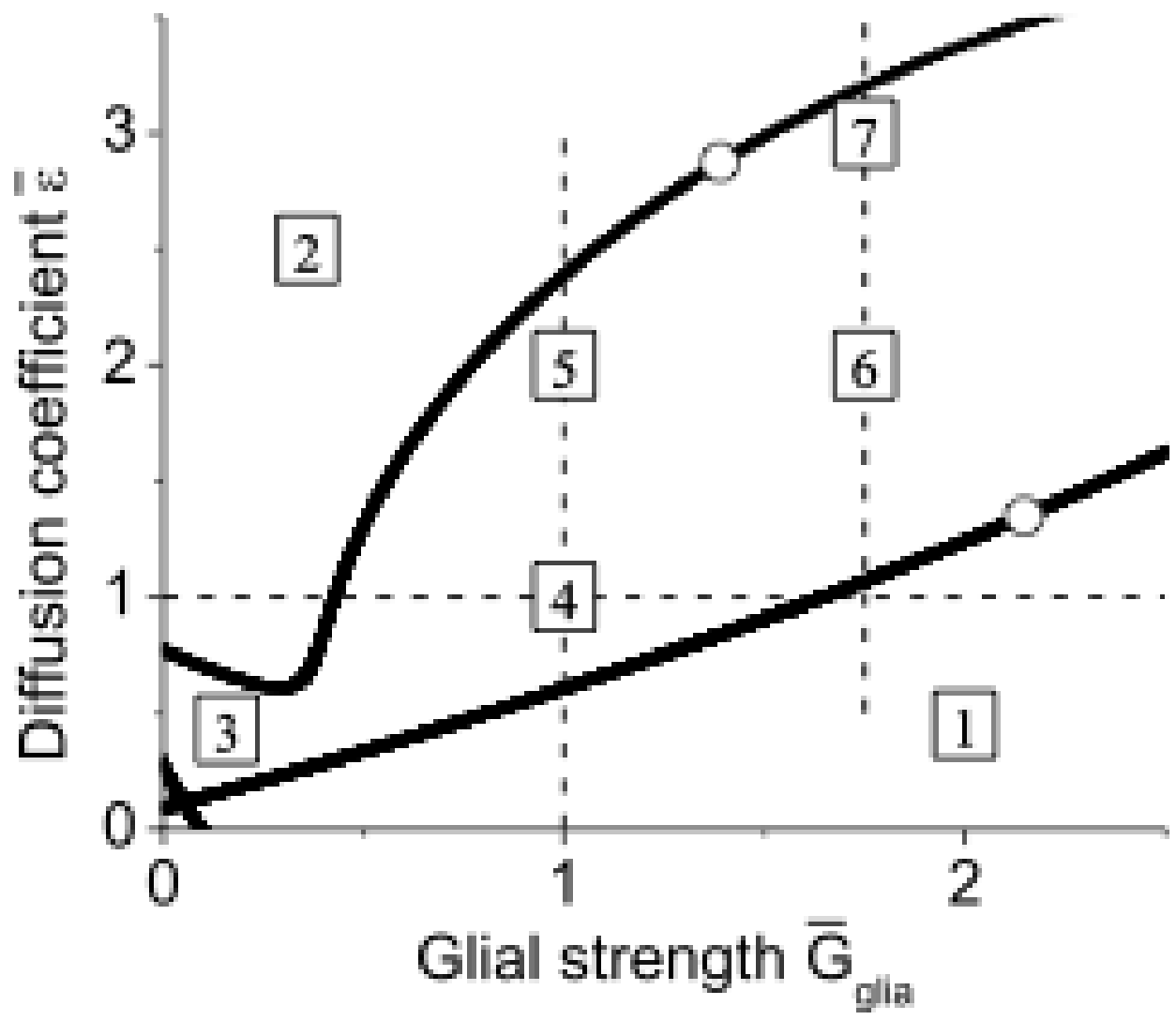

Fig. 4

**FIGURES**

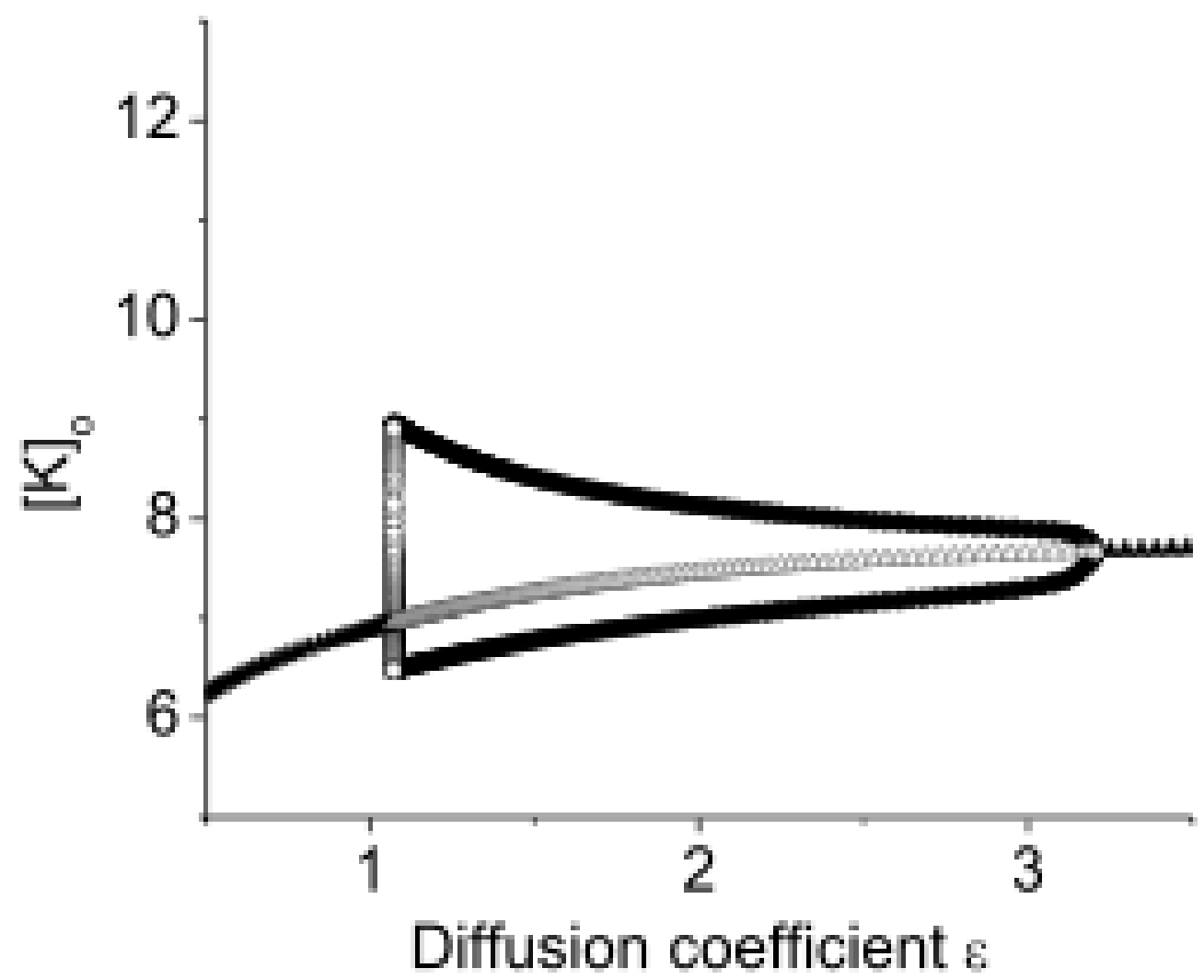

Fig. 5

## FIGURES

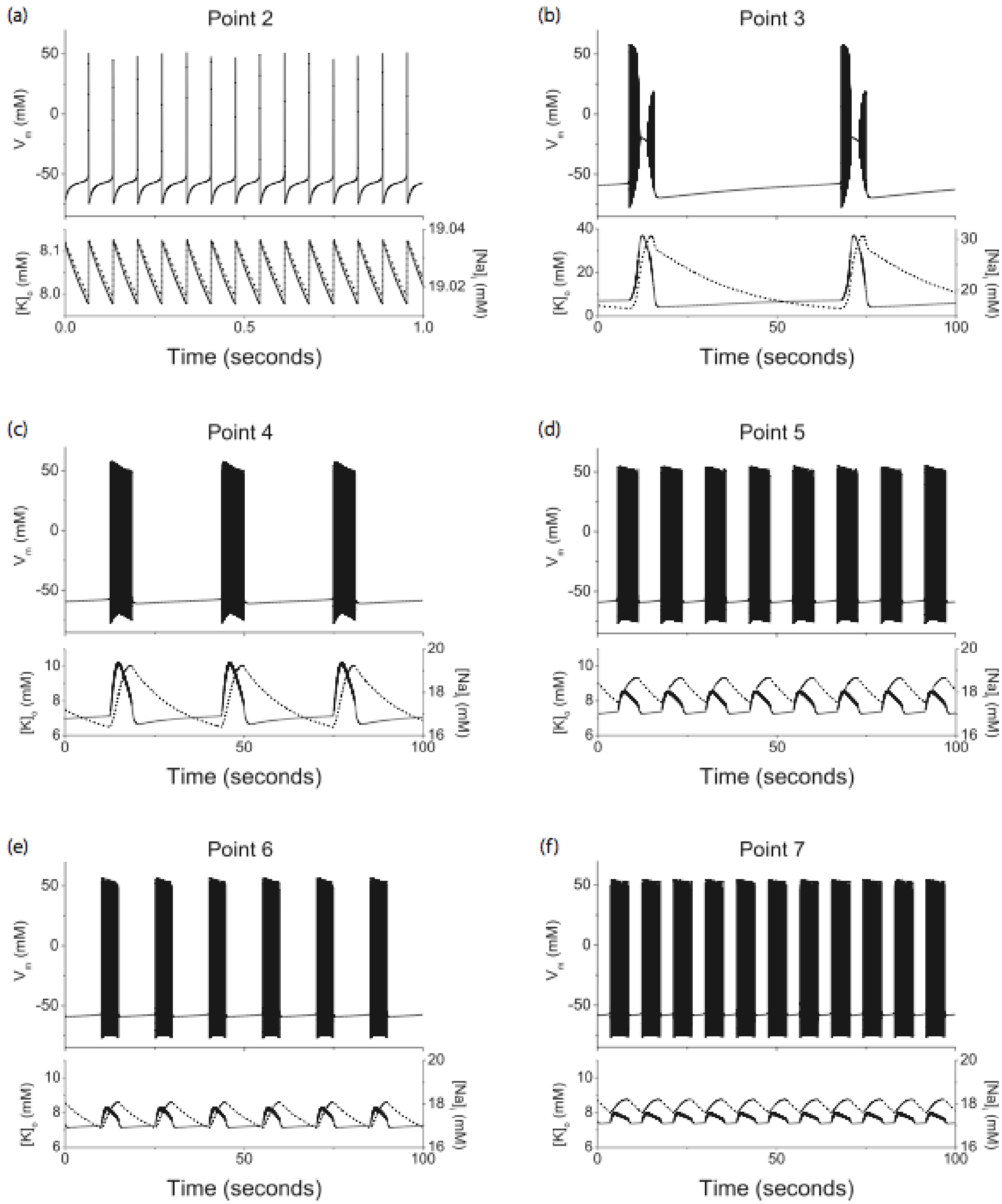

Fig. 6

# FIGURES

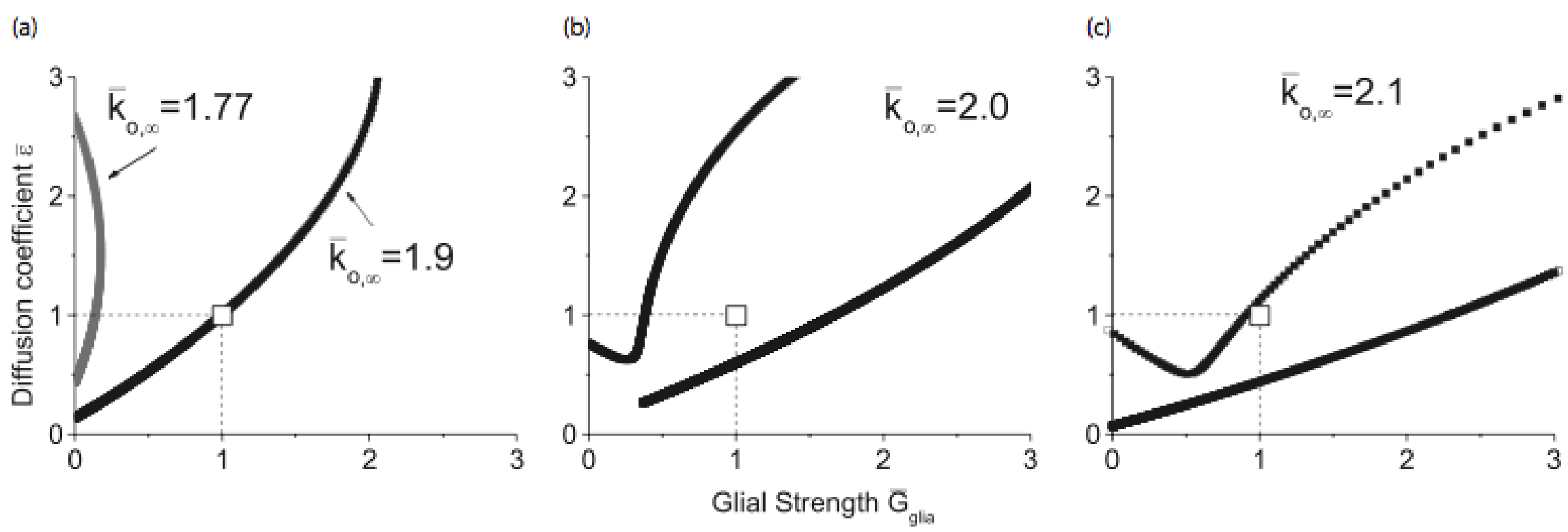

Fig. 7

## FIGURES

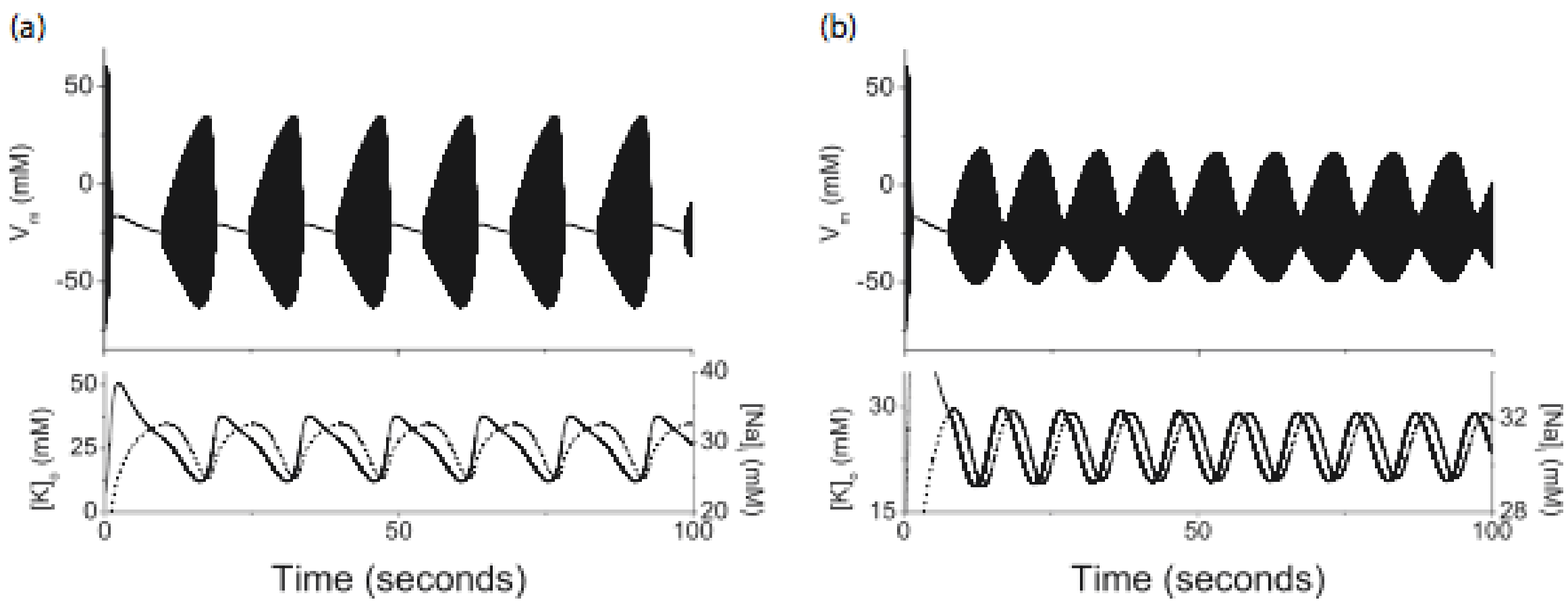

Fig. 8